\documentclass[aps,prl,twocolumn,showpacs,unsortedaddress]{revtex4-1}
\usepackage{epsfig,pstricks,graphicx}
\usepackage{amsmath,amssymb}
\usepackage[mathscr]{eucal}
\usepackage{color}

\def\CC{{\rm\kern.24em \vrule width.04em height1.46ex depth-.07ex
\kern-.30em C}}
\def\RR{{\rm
         \vrule width.04em height1.58ex depth-.0ex
         \kern-.04em R}}
\def\P{{\rm I\kern-.25em P}}
\def\id{{\rm 1\kern-.22em l}}

\newcommand{\bra}[1]{\left\langle #1 \right |}
\newcommand{\ket}[1]{\left | #1 \right\rangle}

\newcommand{\oost}{\frac{1}{\sqrt{2}}}
\newcommand{\rme}{\ensuremath{\mathrm{e}}}
\newcommand{\rmi}{\ensuremath{\mathrm{i}}}
\newcommand{\rmd}{\ensuremath{\mathrm{d}}}

\newcommand{\tr}{\operatorname{tr}}

\begin{document}

\title{Entanglement 
       of three-qubit Greenberger-Horne-Zeilinger-symmetric states
      }
\author{Christopher Eltschka}
\affiliation{Institut f\"ur Theoretische Physik, 
         Universit\"at Regensburg, D-93040 Regensburg, Germany}
\author{Jens Siewert}
\affiliation{Departamento de Qu\'{\i}mica F\'{\i}sica, Universidad del Pa\'{\i}s Vasco -
             Euskal Herriko Unibertsitatea, 48080 Bilbao, Spain}
\affiliation{Ikerbasque, Basque Foundation for Science, 48011 Bilbao, Spain}

\begin{abstract}
The first characterization of mixed-state entanglement was achieved
for two-qubit states in Werner's seminal work
 [Phys.\ Rev.\ A {\bf 40}, 4277  (1989)].
A physically important extension of this result concerns mixtures
of a pure entangled state (such as the Greenberger-Horne-Zeilinger [GHZ] state)
and the completely unpolarized state.
These mixed states serve as benchmark for the robustness of entanglement. 
They share the same symmetries as the GHZ state. 
We call such states GHZ-symmetric.
Despite significant progress
their multipartite entanglement properties have remained an open problem.
Here we give a complete description of the entanglement in 
the family of three-qubit GHZ-symmetric states and, in particular, 
of the three-qubit generalized Werner states.
Our method relies on the appropriate parameterization of the states and
on the invariance of entanglement properties under 
general local operations.  An immediate application of our results is
the definition of a symmetrization
witness for the entanglement class of arbitrary three-qubit states.
\end{abstract}
\pacs{03.67.-a, 03.67.Mn}
\maketitle

{\em Introduction. -- } 
Entanglement is the essential resource for 
many tasks in quantum information
processing~\cite{review1,review2}. Therefore, it is desirable to 
precisely characterize the entanglement contained in a quantum state.
While our understanding of pure-state entanglement has significantly
improved in recent years, entanglement in mixed
states has remained 
a notoriously difficult subject, 
despite numerous impressive results, e.g.,~\cite{Pittenger2000,Acin2001,Werner2001,Eltschka2008,GuehneToth2009,Jung2009,Hiesmayr2010,Guehne2010,Fei2011}. Due to the 
tremendous experimental progress in producing and controlling multi-qubit 
entanglement (e.g., Refs.~\cite{Blatt14ions2011,Pan2007,Weinfurter2009,Martinis2010,Schoelkopf2010}) this has become also a practical problem, as an accurate assessment
of the experimental results is required. The universal tool here 
are entanglement witnesses~\cite{Horodecki1996,Terhal2000,Lewenstein2000,GuehneToth2009}.
Despite its flexibility and success in detecting entanglement and 
distinguishing entanglement classes, characterization by means of
witnesses is not always satisfactory.  
Enhancing the quality of entanglement witnesses requires improvement in
the underlying entanglement theory.

An important and in general unsolved question of both practical and theoretical interest
is how much noise admixture pure-state entanglement can
sustain. Mathematically, this question can be cast as follows.
We consider a pure  state $\ket{\psi_{\mathrm{ME}}}$ of $N$ qubits
that contains a maximum amount of a certain entanglement type. 
This state gets mixed with the operator $\frac{1}{2^N}\id_{2^N}$
which describes the
maximally mixed state of $N$ qubits, serving as a model of unpolarized
noise:
\begin{equation}
  \rho_{\mathrm{WS}}(p) = p\ \ket{\psi_{\mathrm{ME}}}\!\bra{\psi_{\mathrm{ME}}}\ +\ 
                 (1-p)\ \frac{1}{2^N} \id_{2^N}\ \ .
\label{Werner-allg}
\end{equation}
The question then is how small $p\ \ (0\leq p\leq 1)$ 
can be chosen such that $\rho_{\mathrm{WS}}(p)$
still contains a finite amount of the considered entanglement.

For two qubits ($N=2$), one substitutes 
$\ket{\psi_{\mathrm{ME}}}$ with the Bell state
$\ket{\Psi^{-}}=\oost\left(\ket{01}-\ket{10}\right)$. Then, $\rho_{\mathrm{WS}}(p)$ represents
the so-called Werner states
\cite{Werner1989}. Although Werner defined them through the symmetry
under local unitaries $U\otimes U$, the generalizations of Eq.~\eqref{Werner-allg}
to three and more qubits are often termed 
{\em generalized Werner states}~\cite{Pittenger2000,Brylinski2002}. Throughout 
this article, we shall consider three qubits ($N=3$) and the maximally 
entangled GHZ state
\begin{equation}
  \label{eq:GHZ}
  \ket{\mathrm{GHZ}}
           \ =\ 
           \oost\left(\ket{000}\ +\ \ket{111}\right)\ \equiv\ \ket{\mathrm{GHZ}_+}\ \ .
\end{equation}
We shall give a complete characterization of the entanglement in 
\begin{equation}
  \label{eq:rhowerner}
\rho_{\mathrm{WS}}(p)\ =\ p\ket{\mathrm{GHZ}}\!\bra{\mathrm{GHZ}}\ +\ \frac{1-p}{8}\ \id_8 
\end{equation}
and the 
entire family of states with the same symmetry, the GHZ symmetry (see below).
After reviewing the known results
we present a parameterization for this family that allows to 
deduce the entanglement type for any given element of the family.
Finally we show that our findings can be used as a  
witness to detect the entanglement type of arbitrary three-qubit
states. 

For two qubits, Eq.~(\ref{Werner-allg}) gives the
  standard Werner state after
  replacing $\ket{\psi_{\mathrm{ME}}}$ with the Bell state $\ket{\Psi^-}$. 
In the two-qubit case there is only one type of
  entanglement, and therefore the problem reduces to finding the
  maximal value of $p$ such that the state is still not entangled. 
It can easily be found by computing the concurrence~\cite{Wootters1998}.

The three-qubit case,  however, is more complex.
A state can either be completely separable, biseparable or tripartite
entangled. Moreover,
there are two inequivalent classes of tripartite entanglement, the
GHZ type and the $W$ type~\cite{Duer2000}. In the space of density matrices, there is
a hierarchy of entangled states~\cite{Acin2001}: the convex
hull of the $W$-type states includes the true $W$ states, the
biseparable and the separable ones while the set of GHZ-type states contains 
all other classes.

The three-qubit generalized Werner states
$\rho_{\mathrm{WS}}(p)$ are known to
be fully separable if and only if $p\leq p_{\mathrm{sep}}=\frac{1}{5}$~\cite{Pittenger2000} 
and
biseparable if and only if $p\leq p_{\mathrm{bisep}}=\frac{3}{7}$ \cite{Guehne2010}. 
One aim of this
article is to find the value $p_W$ such that $\rho_{\mathrm{WS}}(p)$ is of 
$W$ type for
$p\leq p_W$ and of GHZ type for $p>p_W$. It turns out that it is
advantageous to extend the problem to all mixed states which can be
written as affine combinations of $\ket{\mathrm{GHZ}_+}$, 
$\ket{\mathrm{GHZ}_-} = \oost(\ket{000}-\ket{111})$, 
and the maximally mixed state. 

{\em Parameterization of GHZ-symmetric states. -- }
We solve the problem by exploiting its symmetry.
The GHZ state, and thus also $\rho_{\mathrm{WS}}(p)$, is invariant 
under the following transformations (and combinations thereof):
{\em (i)} qubit permutations, 
{\em (ii)} simultaneous three-qubit flips 
      (i.e., application of  $\sigma_x\otimes\sigma_x\otimes \sigma_x$), 
{\em (iii)} qubit rotations about the $z$ axis of the form
  \begin{equation}
    \label{eq:zrot}
    U(\phi_1,\phi_2) = \rme^{\rmi \phi_1 \sigma_z}\otimes\rme^{\rmi \phi_2 \sigma_z}\otimes\rme^{-\rmi
      (\phi_1+\phi_2) \sigma_z}\ \ .
  \end{equation}
Here, $\sigma_x$ and $\sigma_z$ are Pauli operators.
We refer to the invariance under the operations $(i)-(iii)$
as {\em GHZ symmetry}. Except the qubit
permutations all those operations are local, therefore (and since qubit
permutations always convert GHZ states into GHZ states) 
GHZ symmetry operations will never turn
GHZ-type entanglement into $W$-type entanglement or vice versa.

An important aspect of this symmetry is that for any decomposition of
$\rho_{\mathrm{WS}}(p)$ into pure states there is a GHZ-symmetric decomposition of the 
same entanglement type. 
It is generated by replacing each pure state in the
decomposition with the equal mixture of all states 
obtained from that former state by applying the symmetry operations. 

In order to identify the set of GHZ-symmetric density 
matrices we check
the action of the symmetry operations on its elements $\rho^{\mathrm{S}}$.
First consider the $z$ rotations {\em (iii)}. 
The matrix element $\rho^{\mathrm{S}}_{ijk,lmn}$ is transformed 
by operations according to Eq.~\eqref{eq:zrot} into
$\exp[\rmi(i-k-l+n)\phi_1]\exp[\rmi(j-k-m+n)\phi_2]\rho^{\mathrm{S}}_{ijk,lmn}$.
Since $\phi_1$ and $\phi_2$ can take arbitrary values
  the state remains unchanged only if either the matrix element is zero,
  or
if both $i-k-l+n=0$
and $j-k-m+n=0$. Therefore the only non-zero matrix elements
  are the diagonal elements, $\rho^{\mathrm{S}}_{000,111}$ and $\rho^{\mathrm{S}}_{111,000}$.
Among these  elements,
permutation invariance forces the diagonal elements to 
depend only on the number of 1s in the index. Finally, the invariance
under collective bit flips implies 
$\rho^{\mathrm{S}}_{000,000}=\rho^{\mathrm{S}}_{111,111}$ and 
$\rho^{\mathrm{S}}_{001,001}=\rho^{\mathrm{S}}_{110,110}$. 
Moreover, we have $\rho^{\mathrm{S}}_{000,111}=\rho^{\mathrm{S}}_{111,000}$ 
and thus real off-diagonal matrix elements due to hermiticity.
Given the additional constraint $\tr\rho^{\mathrm{S}}=1$
we find that a state $\rho^{\mathrm{S}}$ is fully specified by 
two independent real parameters. A possible choice is
\begin{eqnarray}
  \label{eq:rhoparams}
 x(\rho^{\mathrm{S}}) &=&\! \frac{1}{2}\!
  \left[\bra{\mathrm{GHZ}_+}\rho^{\mathrm{S}}\ket{\mathrm{GHZ}_+}\! -\! 
        \bra{\mathrm{GHZ}_-}\rho^{\mathrm{S}}\ket{\mathrm{GHZ}_-}\right]\\
 y(\rho^{\mathrm{S}}) &=&\! \frac{1}{\sqrt{3}}
  \left[\bra{\mathrm{GHZ}_+}\rho^{\mathrm{S}}\ket{\mathrm{GHZ}_+}\ +\ \right.
\nonumber
\\ 
      && \ \ \ \ \ \ \ \ \ \ \ \ \ \ \ +\ \left.
        \bra{\mathrm{GHZ}_-}\rho^{\mathrm{S}}\ket{\mathrm{GHZ}_-} - \frac{1}{4}\right]
\end{eqnarray}
such that the Euclidean metric in the
$(x,y)$ plane coincides with the Hilbert-Schmidt metric on the density
matrices. The completely mixed state is located at the origin. 
The set of states $\rho^{\mathrm{S}}$ forms a triangle in the $(x,y)$ plane
(see Fig.~1). 
The generalized Werner states (\ref{eq:rhowerner}) are found on the straight line
  $y=\frac{\sqrt{3}}{2}x$ connecting the origin with the GHZ state. We
  call it the `Werner line'.
\begin{figure}[tbh]
  \centering
  \includegraphics[width=1.0\linewidth]{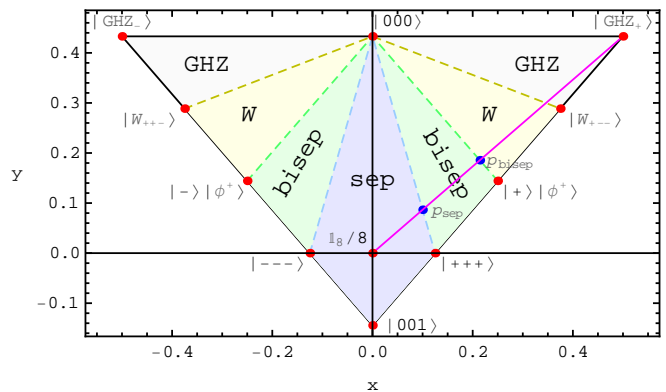}
  \caption{The convex set of GHZ-symmetric density matrices $\rho^{\mathrm{S}}$. 
           The upper corners of the triangle are the standard GHZ state 
    $\ket{\mathrm{GHZ}_+}$, and
    $\ket{\mathrm{GHZ}_-}$. Note that these are the only pure states. 
    Applying $\sigma_z$ to
    any one of the qubits changes the sign of $x$. Therefore for 
    properties invariant under local unitaries, we have a mirror symmetry 
    about the $y$ axis.\\
    At the centre of the upper horizontal line there is the 
    separable state $\frac{1}{2}(\ket{000}\!\bra{000}+\ket{111}\!\bra{111})$.
    The points
    for the pure states $\ket{001}$, 
    $\ket{+\!+\!+}$, $\ket{+}\!\ket{\phi^+}$, $\ket{W_{+--}}$ 
    indicate the positions of the corresponding symmetrized
    mixed state (here we have used the definitions 
    $\ket{\pm}\equiv\oost(\ket{0}\pm\ket{1})$, $\ket{\phi^+}\equiv\oost(\ket{00}+\ket{11})$,
    and $\ket{W_{+--}}\equiv\frac{1}{\sqrt{3}} (\ket{+--}+\ket{-+-}+\ket{--+})$). 
    The solid magenta line (`Werner line') represents the generalized
    Werner states $\rho_{\mathrm{WS}}(p)$ with $p_{\mathrm{sep}}=\frac{1}{5}$ 
    and $p_{\mathrm{bisep}}=\frac{3}{7}$ (see text).
    The two lower dashed lines are  first guesses for the 
    boundaries of fully separable (`sep') and biseparable (`bisep') states from the 
    known values of $p_{\mathrm{sep}}$ and $p_{\mathrm{bisep}}$, respectively. 
    The upper dashed line represents a first guess for the boundary 
    between $W$  and GHZ states 
    as $\ket{W_{+--}}$ is the $W$ state with the largest overlap to 
    $\ket{\mathrm{GHZ}_+}$ 
    (cf.~Ref.~\cite{Acin2001}). The intersection with the Werner
    line occurs at $p=9/13$.
    }
  \label{fig:ghzsymmset}
\end{figure}

For any normalized pure state $\ket{\psi}=(\psi_{000},\ldots,\psi_{111})$, 
there exists a corresponding symmetrized state
\begin{equation}
\label{eq:symmstate}
\rho^{\mathrm{S}}(\psi) = \int\rmd U\, U\ket{\psi}\!\bra{\psi}U^\dagger
\end{equation}
where the integral is understood to cover the entire GHZ symmetry group, 
i.e.,  unitaries $U(\phi_1,\phi_2)$
as in Eq.~\eqref{eq:zrot} and averaging over the discrete symmetries.
The coordinates of the symmetrized state can be inferred 
from the coefficients $\psi_{000}$ and $\psi_{111}$
\begin{eqnarray}
  \label{eq:xpurestate}
  x(\psi) &=& \frac{1}{2}\left(\psi_{000}^*\psi_{111}
                           + \psi_{000}\psi_{111}^*
                         \right)\\
  \label{eq:ypurestate}
  y(\psi) &=&
  \frac{1}{\sqrt{3}}\left(\left|\psi_{000}\right|^2 +
    \left|\psi_{111}\right|^2 -\frac{1}{4}\right)
  \ \ .
\end{eqnarray}
%
{\em Entanglement properties of GHZ-symmetric states. -- }
After finding and suitably parameterizing the set of GHZ-symmetric states 
we want to determine the entanglement class of each state (fully separable,
biseparable, $W$, or GHZ). 
The key idea is that
all states in an entanglement class are equivalent under 
stochastic local operations and classical communication
(SLOCC)~\cite{Duer2000,Bennett2001}. Mathematically, the corresponding
(invertible) local operations 
are represented by the elements of the group $\mathrm{GL}(2,\CC)$. 
That is, applying GL(2,$\CC)$ transformations to any qubit
does not change the entanglement class of a multi-qubit state.

The GHZ-symmetric states of
each SLOCC class form a convex set. We characterize each
set by finding its boundary starting from the separable states.
Our strategy to identify these boundaries is the following.
We fix the $y$ coordinate in the interval
$-1/(4\sqrt{3})\leq y< \sqrt{3}/4$ and then
consider all {\em pure} states $\ket{\psi}$ of the SLOCC class under
consideration whose symmetrized state $\rho^{\mathrm{S}}(\psi)$
has the chosen $y$ according to Eq.~\eqref{eq:ypurestate}.
States at the
boundary are the ones with maximum (or minimum for $x<0$)
$x$ coordinate
according to Eq.~\eqref{eq:xpurestate} for a given $y$, 
termed $x_{\mathrm{max}}$.
Mirror symmetry implies
$x_{\mathrm{min}}=-x_{\mathrm{max}}$, therefore we may restrict our discussion
to $x\geq 0$. 
If $x_{\mathrm{max}}(y)$ does not have the
appropriate curvature the boundary is given
by the convex hull of $x_{\mathrm{max}}(y)$.

We start with an obvious solution that holds for all SLOCC classes.
If, for fixed $y$, the coefficients of the pure state $\ket{\psi}$ can
be chosen equal $|\psi_{000}|^2=|\psi_{111}|^2=\frac{1}{2}(\sqrt{3}y+\frac{1}{4})$ 
the maximum $x$ coordinate is given by 
$x_{\mathrm{max}}=|\psi_{000}||\psi_{111}|=\frac{1}{2}(\sqrt{3}y+\frac{1}{4})$,
i.e., by the lower edge of the triangle of GHZ-symmetric states. 

Now consider the separable pure states $\ket{\psi^{\mathrm{sep}}}$. 
They are equivalent (via
local unitaries) to the state $\ket{000}$
\begin{equation}
   \ket{\psi^{\mathrm{sep}}}\ =\
   \left[
   \bigotimes_{j=1}^{3}
  {\small
   \left( 
    \begin{array}{cc}
      A_j^{\ast} & B_j \\ 
      B_j^{\ast} & -A_j
    \end{array}
    \right)
  }
   \right]
    \ \ket{000}
\end{equation}
where  $|A_j|,|B_j|\leq 1$. For the moduli of the coefficients we find
$|\psi^{\mathrm{sep}}_{000}|=|A_1A_2A_3|$ and 
$|\psi^{\mathrm{sep}}_{111}|=\sqrt{(1-|A_1|^2)(1-|A_2|^2)(1-|A_3|^2)}$. 
Maximizing $x=|\psi^{\mathrm{sep}}_{000}||\psi^{\mathrm{sep}}_{111}|$
subject to the constraint $|\psi^{\mathrm{sep}}_{000}|^2+|\psi^{\mathrm{sep}}_{111}|^2=
\mathrm{const}$ leads to
$   x_{\mathrm{max}} = \left(\frac{1}{4}-\frac{1}{\sqrt{3}}y
                         \right)^{\frac{3}{2}}
$
for $y>0$.
As this function gives a concave boundary (cf.~Fig.~2) we use the convex hull 
\begin{equation}
x_{\mathrm{max}}^{\mathrm{sep}}\ =\ -\frac{\sqrt{3}}{6}y+\frac{1}{8}
\end{equation}
which is identical to the first guess
from the known result 
$p_{\mathrm{sep}}=\frac{1}{5}$~(cf.~Fig.~1).
\begin{figure}[htb]
  \centering
  \includegraphics[width=1.0\linewidth]{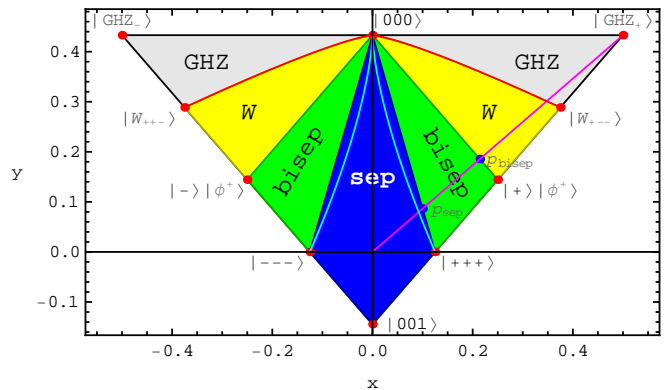}
  \caption{The SLOCC classes of three-qubit GHZ-symmetric states
           $\rho^{\mathrm{S}}$. The dark blue region shows
           the separable states (`sep') with the light blue lines
           $   x = \pm\left(\frac{1}{4}-\frac{1}{\sqrt{3}}y
                         \right)^{\frac{3}{2}}$.
           Green areas represent the biseparable states (`bisep').
           The $W$ states  `$W$' (yellow) and the GHZ states `GHZ' (grey)
           are separated by the curve Eq.~\protect\eqref{b-W}
           (red line). The Werner line (magenta) crosses
           that curve at $p_W\approx 0.6955$.
           Some geometrical aspects are noteworthy. The curve~\eqref{b-W}
           nearly (within a few per cent) describes a circle about the point 
           $\rho^{\mathrm{S}}(001)$. The radius has a minimum in the vicinity
           of the Werner line. Further,
           it is intriguing to note that each SLOCC class
           shares exactly one fourth of the lower edge of the triangle.
            } 
  \label{fig:thefullpicture}
\end{figure}

For biseparable pure states $\ket{\psi^{\mathrm{bisep}}}$
it suffices (due to the subsequent symmetrization)
to consider local equivalence to the state $\ket{0}\!\otimes\!\ket{\phi^{+}}$.
That is, we obtain $\ket{\psi^{\mathrm{bisep}}}$ 
by normalizing the vector
$(G_1\!\otimes G_2\!\otimes G_3)\!\ket{0}\!\otimes\!\ket{\phi^+}$.
Here, 
\[
      G_j\ =\ 
  {\small
   \left( 
    \begin{array}{cc}
      A_j & B_j \\ 
      C_j & D_j
    \end{array}
    \right)
  }\ \ \ ,\ \ j=1,2,3
\]
denotes an arbitrary GL($2,\CC)$ transformation. 
The discussion can be restricted
to $G_2=G_3$ because for
any $\ket{\psi^{\mathrm{bisep}}}$
the two-qubit part can be made permutation symmetric
by a diagonal GL($2,\CC)^{\otimes 2}$ operation 
without
decreasing the coordinates $x(\psi^{\mathrm{bisep}})$,
                       $y(\psi^{\mathrm{bisep}})$
of the corresponding 
symmetrized state.
Maximizing $x$ as before yields 
\begin{equation}
x_{\mathrm{max}}^{\mathrm{bisep}}\ =\ -\frac{\sqrt{3}}{2}y+\frac{3}{8}
\end{equation}
for $y>\frac{1}{4\sqrt{3}}$.
Again this boundary coincides with the one inferred from 
$p^{\mathrm{bisep}}=\frac{3}{7}$ (see Fig.~1).

The general pure $W$ state $\ket{\psi^W}$ is found by normalizing
$(G_1\!\otimes G_2\!\otimes G_3)\!(\ket{001}+\ket{010}+\ket{100})$.
In analogy with the separable states, maximization of 
$x=|\psi^W_{000}||\psi^W_{111}|$ subject to the
constraint $|\psi^{W}_{000}|^2+|\psi^{W}_{111}|^2=
{\mathrm{const}}$ shows that the maximum is reached for
$G_1=G_2=G_3$.  It leads to polynomial equations whose solutions
are given, for convenience,  in parameterized
form (with $0\leq v\leq 1$)
\begin{equation}
    x^W_{\mathrm{max}}=\frac{v^5+8v^3}{8(4-v^2)}\ \ \ ,\ \ \
    y=\frac{\sqrt{3}}{4}\frac{4-v^2-v^4}{4-v^2}
\label{b-W}
\end{equation}
where $y\geq\frac{1}{2\sqrt{3}}$. The second derivative of 
$x_{\mathrm{max}}^{W}(y)$ shows that the boundary is indeed convex.
This completes the characterization of SLOCC classes for GHZ-symmetric
three-qubit states.

A particularly interesting point is the intersection of the curve~\eqref{b-W}
with the Werner line $y_{\mathrm{WS}}=\frac{\sqrt{3}}{2}x$. The corresponding
parameter $v_W$ solves the equation
\[
            1\ =\ 4\frac{4-v_W^2-v_W^4}
                        {v_W^3(v_W^2+8)}
\]
such that $p_W=0.6955427\ldots$

{\em Symmetrization witness. -- }
Although these results might seem of purely theoretical interest
they have a surprising application for arbitrary three-qubit
states. Suppose $\rho$ is such a state. The twirling operation
in Eq.~\eqref{eq:symmstate} generates the corresponding symmetrized
state $\rho^{\mathrm{S}}(\rho)$.
The SLOCC class  of $\rho$ cannot be lower  in the hierarchy
  described in the introduction than that of
$\rho^{\mathrm{S}}(\rho)$. 
For example, a $W$ state can be projected  by the twirling
operation Eq.~\eqref{eq:symmstate}
onto a $W$ state, a biseparable state or a fully separable state,
  but not onto a GHZ state. 
Hence, 
the GHZ-symmetrized state $\rho^{\mathrm{S}}(\rho)$ can be used to witness the SLOCC
class of the original state $\rho$, simply by reading off the coordinates of
 $\rho^{\mathrm{S}}(\rho)$ in Fig.~2.
These coordinates $x(\rho)$ and $y(\rho)$ are obtained
from the matrix elements of $\rho$:
\begin{eqnarray}
    x(\rho)\ &=&\ \frac{1}{2} \left( \rho_{000,111}+\rho_{111,000}  \right)
\nonumber\\
    y(\rho)\ &=&\ \frac{1}{\sqrt{3}}\left(\rho_{000,000}+\rho_{111,111}-\frac{1}{4}\right)
    \ \ .
\nonumber
\end{eqnarray}
We will discuss the optimization of this method elsewhere.

Summarizing, we have determined exactly the entanglement properties
of an entire family of high-rank (mostly eight) 
mixed three-qubit states with the same symmetry as the
GHZ state. In particular, we have solved the problem for the three-qubit
generalized Werner state which is a reference for 
multi-qubit mixed-state entanglement. A practically relevant application
of this result is a simple method for detecting the SLOCC
class  of arbitrary three-qubit states. 

{\em Acknowledgements. -- }
This work was funded by the German Research Foundation within SFB 631 
and SPP 1386 (C.E.), and by Basque Government grant IT-472 (J.S.).
The authors thank J.\ Fabian and K.\ Richter 
for their support.
%
%
%
%

%


\begin{thebibliography}{99}
%
\bibitem{Werner1989} 
R.F.\ Werner, 
    Phys.\ Rev.\ A {\bf 40}, 4277  (1989).
%
\bibitem{review1}
M.B.\ Plenio  and S.\ Virmani, 
Quant.\ Inf.\ Comput.\ {\bf 7}, 1 (2007).
\bibitem{review2}
R.\ Horodecki, P.\ Horodecki, M.\ Horodecki, and K.\ Horodecki, 
 Rev.\ Mod.\ Phys.\ {\bf 81}, 865 (2009).
%
\bibitem{Pittenger2000} 
A.O.\  Pittenger  and M.H.\ Rubin, 
  Optics Comm.\ {\bf 179}, 447 (2000);
W.\ D\"ur and J.I.\ Cirac,
Phys.\ Rev.\ A {\bf 61}, 042314 (2000).
%
\bibitem{Acin2001}
A.\ Acin, D.\ Bru{\ss}, M.\ Lewenstein, and A.\ Sanpera, 
Phys.\ Rev.\ Lett.\ {\bf 87}, 040401 (2001).
%
\bibitem{Werner2001} 
T.\ Eggeling and R.F.\ Werner, 
    Phys.\ Rev.\ A {\bf 63}, 042111  (2001).
%
\bibitem{Eltschka2008}
C.\ Eltschka, A.\ Osterloh, J.\ Siewert, and A.\ Uhlmann, 
New J.\ Phys.\ {\bf 10}, 043014 (2008).
%
\bibitem{GuehneToth2009}
O.\ G{\"u}hne and G.\  Toth, 
Phys.\ Rep.\ {\bf 474}, 1 (2009).
%
\bibitem{Jung2009} 
E.\ Jung, M.R.\ Hwang, D.\ Park, and J.W.\ Son,
Phys.\ Rev.\ A {\bf 79}, 024306 (2009).
%
\bibitem{Guehne2010}
O.\ G{\"u}hne and M.\ Seevinck, 
New J.\ Phys.\ {\bf 12}, 053002 (2010).
%
\bibitem{Hiesmayr2010}
M.\ Huber, F.\ Mintert, A.\ Gabriel, and B.C.\ Hiesmayr, 
Phys.\ Rev.\ Lett.\ {\bf 104}, 210501 (2010).
%
\bibitem{Fei2011} 
S.J.\ He, X.H.\ Wang, S.M.\ Fei, H.X.\ Sun, and Q.Y.\ Wen,
Comm.\ Theoret.\ Phys.\ {bf 55}, 251 (2011).
%
\bibitem{Blatt14ions2011} 
   T.\ Monz, P.\ Schindler, J.T.\ Barreiro, M.\ Chwalla, 
   D.\ Nigg, W.A.\ Coish, M.\  Harlander, W.\ Hansel, 
   M.\ Hennrich, and R.\ Blatt, 
Phys.\ Rev.\ Lett.\ {\bf 106}, 130506 (2011).
%
\bibitem{Pan2007} 
  C.-Y.\ Lu, X.-Q.\ Zhou, O.\ G{\"u}hne, W.-B.\ Gao, J.\ Zhang, Z.-S.\ Yuan,
  A.\ Goebel, T.\ Yang, and J.-W.\ Pan,
Nat.\ Phys.\ {\bf 3}, 91 (2007).
%
\bibitem{Weinfurter2009} 
W.\ Wieczorek, R.\ Krischek, N.\ Kiesel, P.\ Michelberger, G.\ Toth, 
and H.\ Weinfurter, 
Phys.\ Rev.\ Lett.\ {\bf 103}, 020504 (2009).
%
\bibitem{Martinis2010} 
M.\ Neeley, R.C.\ Bialczak, M.\ Lenander, E.\ Lucero, 
M.\ Mariantoni, A.D.\ O'Connell, D.\  Sank, H.\  Wang, M.\  Weides, 
J.\ Wenner, 
Y.\ Yin, T.\ Yamamoto, A.N.\ Cleland, and J.M.\ Martinis, 
Nature {\bf 467}, 570 (2010).
%
\bibitem{Schoelkopf2010} 
L.\ DiCarlo, M.D.\ Reed, L.\ Sun, B.R.\ Johnson, J.M.\ Chow, J.M.\ Gambetta, 
L.\ Frunzio, S.M.\ Girvin, M.H.\ Devoret, and R.J.\ Schoelkopf,
Nature {\bf 467}, 574 (2010).
%
\bibitem{Horodecki1996} 
M.\ Horodecki, P.\ Horodecki, and R.\ Horodecki, 
Phys.\ Lett.\ A {\bf 223}, 1 (1996).
%
\bibitem{Terhal2000} 
B.M.\ Terhal, 
Phys.\ Lett.\ A {\bf 271}, 319 (2000).
%
\bibitem{Lewenstein2000}
M.\ Lewenstein, B.\ Kraus, J.I.\ Cirac, and P.\ Horodecki, 
 Phys.\ Rev.\ A {\bf 62}, 052310 (2000).
%
\bibitem{Brylinski2002}
   B.-G.\ Englert,  and N.\ Metwally, 
   in
   {\em Mathematics 
    of Quantum Computation}, edited by G. Chen and 
    R.-K. Brylinski
   (Chapman and Hall/CRC Press, London, 2002), Chap. 2. 
%
\bibitem{Wootters1998} 
   W.K.\ Wootters, 
   Phys.\ Rev.\ Lett.\ \textbf{80}, 2245 (1998).
%
\bibitem{Duer2000} 
 W.\ D\"ur, G.\ Vidal, and J.I.\ Cirac,
   Phys.\ Rev.\ A {\bf 62}, 062314 (2000).
%
\bibitem{Bennett2001}
 C.H.\ Bennett,  S.\ Popescu, D.\ Rohrlich, J.A.\ Smolin, and
   A.V.\ Thapliyal, 
   Phys.\ Rev.\ A {\bf 63}, 012307 (2001).
%
%

%
\end{thebibliography}
\end{document}